\documentclass[conference]{IEEEtran}
\IEEEoverridecommandlockouts
\usepackage{cite}
\usepackage{amsmath,amssymb,amsfonts}
\usepackage{gensymb}
\usepackage[hidelinks]{hyperref}
\usepackage[export]{adjustbox}
\usepackage{multirow}
\usepackage{graphicx}
\usepackage{textcomp}
\usepackage{xcolor}
\usepackage{comment}
\usepackage{subfigure}
\usepackage[ruled,vlined]{algorithm2e}
\usepackage{algpseudocode}

\def\BibTeX{{\rm B\kern-.05em{\sc i\kern-.025em b}\kern-.08em
    T\kern-.1667em\lower.7ex\hbox{E}\kern-.125emX}}
\begin{document}

\title{Perception of Personality Traits in Crowds of Virtual Humans
}


\author{
\IEEEauthorblockN{Lucas Nardino}
\IEEEauthorblockA{\textit{Virtual Humans Lab} \\
\textit{School of Technology} \\
\textit{Pontifical Catholic University} \\
\textit{of Rio Grande do Sul} \\Porto Alegre, Brazil \\lucas.nardino@edu.pucrs.br}
\\
\IEEEauthorblockN{Enzo Krzmienszki}
\IEEEauthorblockA{\textit{Virtual Humans Lab} \\
\textit{School of Technology} \\
\textit{Pontifical Catholic University} \\
\textit{of Rio Grande do Sul} \\Porto Alegre, Brazil \\enzo.neves@edu.pucrs.br}
\\
\IEEEauthorblockN{Vinícius Jurinic Cassol}
\IEEEauthorblockA{\textit{Virtual Humans Lab} \\
\textit{School of Technology} \\
\textit{Pontifical Catholic University} \\
\textit{of Rio Grande do Sul} \\Porto Alegre, Brazil \\vinicius.cassol@edu.pucrs.br}

\and
\IEEEauthorblockN{Diogo Schaffer}
\IEEEauthorblockA{\textit{Virtual Humans Lab} \\
\textit{School of Technology} \\
\textit{Pontifical Catholic University} \\
\textit{of Rio Grande do Sul} \\Porto Alegre, Brazil \\diogo.schaffer@acad.pucrs.br}
\\
\IEEEauthorblockN{Victor Flávio de Andrade Araujo}
\IEEEauthorblockA{\textit{Virtual Humans Lab} \\
\textit{School of Technology} \\
\textit{Pontifical Catholic University} \\
\textit{of Rio Grande do Sul} \\Porto Alegre, Brazil \\victor.flavio@acad.pucrs.br}
\\
\IEEEauthorblockN{Rodolfo Migon Favaretto}
\IEEEauthorblockA{\textit{Virtual Humans Lab} \\
\textit{School of Technology} \\
\textit{Pontifical Catholic University} \\
\textit{of Rio Grande do Sul} \\Porto Alegre, Brazil \\	rodolfo.favaretto@gmail.com}

\and
\IEEEauthorblockN{Felipe Elsner}
\IEEEauthorblockA{\textit{Virtual Humans Lab} \\
\textit{School of Technology} \\
\textit{Pontifical Catholic University} \\
\textit{of Rio Grande do Sul} \\Porto Alegre, Brazil \\f.elsner@edu.pucrs.br}
\\
\IEEEauthorblockN{Gabriel Fonseca Silva}
\IEEEauthorblockA{\textit{Virtual Humans Lab} \\
\textit{School of Technology} \\
\textit{Pontifical Catholic University} \\
\textit{of Rio Grande do Sul} \\Porto Alegre, Brazil \\gabriel.fonseca94@edu.pucrs.br}
\\
\IEEEauthorblockN{Soraia Raupp Musse}
\IEEEauthorblockA{\textit{Virtual Humans Lab} \\
\textit{School of Technology} \\
\textit{Pontifical Catholic University} \\
\textit{of Rio Grande do Sul} \\Porto Alegre, Brazil \\soraia.musse@pucrs.br}
}

\newcommand\red[1]{{#1}}

\maketitle

\begin{abstract}
This paper proposes a perceptual visual analysis regarding the personality of virtual humans. Many studies have presented findings regarding the way human beings perceive virtual humans with respect to their faces, body animation, motion in the virtual environment and etc. 
\red{We are interested in investigating the way people perceive visual manifestations of virtual humans' personality traits when they are interactive and organized in groups.}
Many applications in games and movies can benefit from the findings regarding the perceptual analysis with the main goal to provide more realistic characters and improve the users' experience. We provide experiments with subjects and obtained results indicate that, although is very subtle, people perceive more the \red{extraversion (the personality trait that we measured)}, into the crowds of virtual humans, when interacting with virtual humans behaviors, than when just observing as a spectator camera.
\end{abstract}


\section{Introduction}
\label{sec:introduction}

\footnote{Draft version made for arXiv: \url{https://arxiv.org/}}Since the pioneer work proposed by Thalmann and Musse~\cite{MusseThalmann1997}, many other methods were proposed for crowd simulation, each one with a significant contribution. There are methods that deal with crowds from a microscopic point of view~\cite{Pelechano:2007,Reynolds:1987}, as well methods that deal with a macroscopic point of view~\cite{hughes2002continuum,Treuille:2006}, and, even, methods that combine both microscopic and macroscopic simulation strategies~\cite{antonitsch2019bioclouds}. Others explored how to compare crowds~\cite{Musse2012}, high dense crowds~\cite{Pelechano:2007,Narain:2009}, heterogeneous behaviors~\cite{Zheng:2016}, navigation control~\cite{Paris2007}, and personality traits for agents\cite{durupinar2009ocean,knob2018simulating,silva2020fastForward}.

Despite the great number of methods proposed for the most varied range of subjects concerning crowd simulation, only very few of them tackled the problem of perceptual analysis of behaviors in crowds.
Indeed, human perception is essential for Computer Graphics (CG). Several techniques developed in the past were based on knowledge of human vision, for example, the interpretation of visual stimuli~\cite{zell2019perception}. These stimuli generate information, which is processed and placed in a specific context. Human perception is a theme present in several researches in CG~\cite{zell2015stylize, zell2019perception}, and it is considered very relevant when discussing the evolution of virtual humans. Virtual humans can be observed through stimuli such as images, videos, games, and Virtual Reality interactions, among others. For these virtual humans do not generate uncomfortable perceptions and falling into an Uncanny Valley~\cite{mori1970bukimi}, they need to present characteristics common to human beings, such as emotions, personality traits, interactions, expressions, etc.

Crowd perception is very important for learning about group behavior, in which observers can see interpersonal interaction on a collective level~\cite{lamer2018rapid}. The area of crowd perception has grown in recent years in several scientific researches (both through psychology and computing), such as perception of different models of agents in crowd simulations~\cite{mcdonnell2008clone}, perception of geometric and cultural features in virtual crowds~\cite{araujo2019much, araujo-cgi:2021}, perception of density in virtual crowds from two points of view~\cite{Yang:2018}, effects on users during interaction with a virtual crowd in an immersive virtual reality environment~\cite{volonte:2020}, studies of social categorization and emotions in crowds using ensemble coding~\cite{lamer2018rapid, goldenberg2021crowd}, among other researches. However, these methods do not focus on the perception of interactions between agents, between agent and user, and the impact of geometric personalities and emotions (that is, no facial and body expressions) on the perception of these interactions. In this work, we define three hypotheses we want to answer: 

\begin{itemize}
\item $H0_1$ defining that people with only observational control of agents in the crowd (do not interfere with crowd dynamics) perceive interactions similarly to people with control of agents in the crowd (the user is considered a crowd agent);

\item $H0_2$ defining that people with only observational control of crowd agents perceive different personalities and emotions similarly to people with control of crowd agents. \red{In this case, as in our work we only use extraversion personality trait, different personalities mean that an agent can or cannot be extraverted}; 

\item $H0_3$ defining that the perception of interactions in crowds is not related to the perception of different personalities and emotions;

\end{itemize}

To try to answer the hypotheses, we created three scenarios with virtual crowds: \textit{i)} Scenario 1, in which a user controlled a first-person camera throughout the entire scenario, without interfering with the agents' behavior; \textit{ii)} Scenario 2, in which a user also controlled a first-person camera throughout the entire scenario, but he/she is considered as one agent of the simulated crowd, using the BioCrowds~\cite{de2012simulating} model; 
\textit{iii)} Scenario 3, in which a user is also an agent in the crowd, but the simulated crowd is different from Scenario 2 because we use an extension of BioCrowds model, called Normal Life behaviors~\cite{Rockenbach2018}, i.e., people are not in emergent situation. As the contribution of this paper, we introduced in BioCrowds the Extraversion factor to be distributed among the agents, so they are impacted by their levels of extraversion when applying their motion. 
Such factor is inspired in the personality traits methods, as proposed by Durupinar et al.~\cite{durupinar2009ocean}. 
From the observations and interactions with the scenarios, people answered questions about how they perceive the agents' interactions, and their different personalities and emotions, as discussed in this paper. 


This paper is organized as follows: Section~\ref{sec:related_work} presents the related work, while Section~\ref{sec:proposed_model} presents the methodology proposed. Section~\ref{sec:results} presents the results achieved with our method and evaluation with subjects. Finally, Section~\ref{sec:conclusion} presents the final considerations and future work of our method.

\section{Related Work}
\label{sec:related_work}


This section discusses some work related to pedestrian and crowd behavioral analysis, focusing on personality traits, emotion, and perception. Knob et al.~\cite{Knob:2018} presented work related to visualizing interactions between pedestrians in video sequences and virtual agents in crowd simulations. OCEAN-based factors gave interactions for each pedestrian and agent. OCEAN~\cite{Digman:1990, John:1990} is the most commonly used personality trait model for this type of analysis, also known as the Big-Five: \textit{O - Openness to experience}: ``the active seeking and appreciation of new experiences''; \textit{C - Conscientiousness}: ``degree of organization, persistence, control, and motivation in goal directed-behavior''; \textit{E - Extraversion}: ``quantity and intensity of energy directed outwards in the social world''; \textit{A - Agreeableness}: ``the kinds of interaction an individual prefers from compassion to tough-mindedness"; \textit{N - Neuroticism}: ``how much prone to psychological distress the individual is''~\cite{Lordw:2007}. Durupinar et al.~\cite{Durupinar:2016} also used OCEAN to visually represent personality traits.

Visual representation of agents is given in various ways, for example, the animations of agents are based on these two cultural features (OCEAN and emotion). If an agent is sad, the animation will represent that emotion. Yang et al.~\cite{Yang:2018} conducted a study analyzing perception to determine the impact of groups at various densities, using two points of view: top and first-person. In addition to that, they examined what kind of camera position might be best for density perception.


Regarding realism perception, the work proposed by Araujo et al.~\cite{araujo2021perceived} investigated people's perception of characters created using CG, comparing if they feel more comfortable with more recent CG characters or older ones. The authors found out that the perceived comfort about newer CG characters was more significant than the perception of comfort about older CG characters. Also, people's perception of comfort in 2020 was greater than people's perception in 2012.

In another work~\cite{araujo-cgi:2021, araujo2019much}, the authors evaluate the human perception regarding geometric features, personalities, and emotions in avatars. Results indicate that, even without explaining to the participants the concepts of cultural features and how they were calculated (considering the geometric features), in most cases, the participants perceived the personality and emotion expressed by avatars, even without faces and body expressions.

The work proposed by Volonte et al.~\cite{volonte:2020} examined the effects on users during interaction with a virtual human crowd in an immersive virtual reality environment. They found that the users’ were able to interpret the verbal and non-verbal behaviors of the virtual human characters where Positive emotional crowds elicit the highest scores in the variables related to interaction with the virtual characters.

Next, we present the proposed model to generate virtual agents with personality traits \red{(in this case, we just used the extraversion personality trait)} and how we evaluate the people's perception.


\section{Proposed Model}
\label{sec:proposed_model}

This section describes our model to provide agents endowed with personalities, in order to allow the simulation of realistic individuals. Firstly, in Section~\ref{sec:BioCrowds} we briefly describe BioCrowds~\cite{de2012simulating}, in Section~\ref{sec:BioExtension} we describe the Normal Life~\cite{Rockenbach2018} BioCrowds extension 
and finally, in Section~\ref{sec:proposed_model_personality} we detail the personality model.

\subsection{BioCrowds Model}
\label{sec:BioCrowds}

\red{BioCrowds~\cite{de2012simulating} is a model for crowd simulations based on a space colonization algorithm designed to generate leaf venation patterns. In this model, a discrete space is populated by a set of marker points. Virtual agents compete for these markers based on a proximity criterion and capture range, effectively competing for the space in which they occupy and move.} Indeed, each agent $i$ accesses the markers inside its personal space $R_i$ to search for markers that are closest to $i$ than any other agent $j$. So, a marker is only available to the closest agent.


For a given agent $i$, with a set of $N$ available markers $S = \{a_1,a_2, \cdots, a_N\}$, we calculate it's movement vector $\vec{m}$ using Equation~\ref{eq:biocrowds_motion}:

\begin{equation}
\vec{m} = \sum_{k=1}^N w_k (\vec{a}_k - \vec{x}),
\label{eq:biocrowds_motion}
\end{equation}
where $\vec{a}_k$ is the marker's position and $\vec{x}$ is the agent's position. $w_k$ is that marker's weight, calculated from Equation~\ref{eq:biocrowds_marker_weight}:

\begin{equation}
w_k = \frac{f(\vec{g} - \vec{x}, \vec{a}_k - \vec{x} )}{\sum_{l=1}^N f(\vec{g} - \vec{x}, \vec{a}_l - \vec{x} )},
\label{eq:biocrowds_marker_weight}
\end{equation}
where $\vec{g}$ is the position of agent $i$ goal.

To determine function f, let us first assume that all markers $\vec{a}_k$ affecting agent $i$ are at the same distance $\vec{a}_k - \vec{x}$ from this agent. Such function should prioritize markers that lead the agent directly to its goal, i.e., it should (i) reach its maximum when the (nondirected)
angle $\theta$ between $\vec{g} - \vec{x}$ and $\vec{a}_k - \vec{x}$ is equal to $0$\degree; (ii) reach its minimum when $\theta=180$\degree; and (iii) decrease monotonically as $\theta$ increases from $0$ to $180$\degree. Also, if the distances $\vec{a}_k - \vec{x}$ differ, the markers further from the agent should have relatively smaller
weights, to prevent them from dominating the computation of the tentative motion vector $\vec{m}$.
A possible choice for $f$ that satisfies these assumptions is defined in Equation~\ref{eq:biocrowds_formulaF}:
 
\begin{equation}
f(x, y) = \frac{1+cos\theta}{1+||y||},
\label{eq:biocrowds_formulaF}
\end{equation}
where $\theta$ is the angle between $x$ and $y$. Please refer to BioCrowds original paper~\cite{de2012simulating} for further details about the method.

The weights will cause the agent to move towards its goal as long as there are markers available along the way. An agent's movement will be blocked by the absence of markers.

\subsection{BioCrowds Normal Life}
\label{sec:BioExtension}

Helbing et al.~\cite{Helbing2001} present some of the main characteristics of people in normal life evacuations:

\begin{itemize}
\item  In general, pedestrians take into account detours as well as the comfort of walking, thereby minimizing the effort to reach their destination; 
\item Pedestrians prefer to walk with an individual desired speed, which corresponds to the most comfortable walking speed as long as it is not necessary to go faster in order to reach the destination in time;
\item Pedestrians keep a certain distance from other pedestrians and borders. 
\end{itemize}

Using BioCrowds, Rockenbach et al.~\cite{Rockenbach2018} proposed an extension to provide crowds that achieve the main characteristics of normal life~\cite{Helbing2001}.
In this case, the Normal Life behavior aims to improve the realism of agents' behaviors in evacuation scenarios. If we imagine that agents want to evacuate the environment, but without stress, i.e., it is not a panic situation, people will apply some behaviors that are different from the ones applied during a hazardous scenario. 
In this model, the authors proposed the term comfort ($c$) as a function of the available area for each agent. According to Helbing et al.~\cite{Helbing2001}, this area is smaller the more a pedestrian is in a hurry, and still decreases with higher pedestrian density. As proposed in previous work~\cite{Rockenbach2018}, in our method, the sense of personal area was adapted to the number of markers $N_i$ each agent $i$ has. 
So, $c_i$ is defined as a function of the number of available markers (the set $S_i$) a certain agent $i$ has. If the number of markers $N_i$ decreases, then $c_i$ decreases too. So, the agent will gradually shift its focus from its designated goal to looking for a more comfortable space i.e., with more available markers. Actually, we normalize $N_i$ dividing by the maximum number of markers $M$ (empirically defined as $70$, once it is impacted by the world configurations).

With this definition, the comfort factor is in the interval $[0;1]$ for agent $i$, according to Equation~\ref{eq:confort_factor}:
\begin{equation}
c_i = \frac{N_i}{M}.
\label{eq:confort_factor}
\end{equation}
The original BioCrowds~\cite{de2012simulating} model computes the weight of each marker, as defined in Equation~\ref{eq:biocrowds_marker_weight}, by comparing the angle difference between the direction defined from the agent towards its goals and all available markers. In Normal Life BioCrowds~\cite{Rockenbach2018}, the markers weights are computed in order to endow agents with the previously described behavior, i.e to look for a more comfortable space. The new weight affected by comfort ($w_k'$) for agent $i$ is defined by Equation~\ref{eq:confort_weight}:

\begin{equation}
w_{k,i}' = \delta_i.w_{k,i} + (1-\delta_i),
\label{eq:confort_weight}
\end{equation}
where $w_{k,i}$ is the original weight calculated by BioCrowds in Equation~\ref{eq:biocrowds_marker_weight} and $\delta_i$ is the comfort bias for agent $i$ defined by Equation~\ref{eq:confort_weight_factor}:

\begin{equation}
\delta_i = \sin(c_i.\frac{\pi}{2}).
\label{eq:confort_weight_factor}
\end{equation}

Related to Equation~\ref{eq:confort_weight}, agents behave according to original BioCrowds when $\delta_i=1$, i.e., markers weights vary according to the goal direction.  However, when the number of markers decreases, the bias decreases as well, resulting in their weights being more similar, causing the agent to go towards the available markers, even if those do not lead to the goal.  

While crowd behaviors are studied in various scenarios, it is acceptable that various "normal life scenarios" can be different in real life. One possibility is that the crowd is affected by the personality traits of membership and not only responsive to the space around the subjects. This is the main goal of our work and the methodology to achieve that is presented in the next section.

\subsection{Extraversion Personality Trait}
\label{sec:proposed_model_personality}

In order to include personality traits in our agents, we chose the OCEAN (Openness to experience, Conscientiousness, Extraversion, Agreeableness, Neuroticism) psychological traits model, proposed by Goldberg~\cite{goldberg1990alternative}, once it is the most accepted model to define the personality of a person.

In this work, we focused on modeling the Extraversion trait, which reflects the sociability and talkativeness but also, in the geometric sense, how comfortable the individual is around crowds and other groups~\cite{favaretto2017using}. \red{So, the Extraversion trait can affect the comfort of an agent when interacting with a user's avatar.} The Normal Life model dictates how much the agents value their personal space in comparison to the desire to reach their goals. We propose an Extraversion factor $E_i$, for agent $i$, which influences the generated behaviors according to each agent's personality, as to vary how comfortable the agent is with a crowded personal space. We can see in Equation \ref{eq:extraversion} the modified Normal Life Equation \ref{eq:confort_weight}, including the Extraversion factor included.

\begin{equation}
w_{k,i}'' = \delta_i.w_{k,i}.E_i + (1-\delta_i) . ( 1 - E_i).
\label{eq:extraversion}
\end{equation}

\begin{figure}[t!]
  \centering
  \includegraphics[scale=1.0, width=\linewidth]{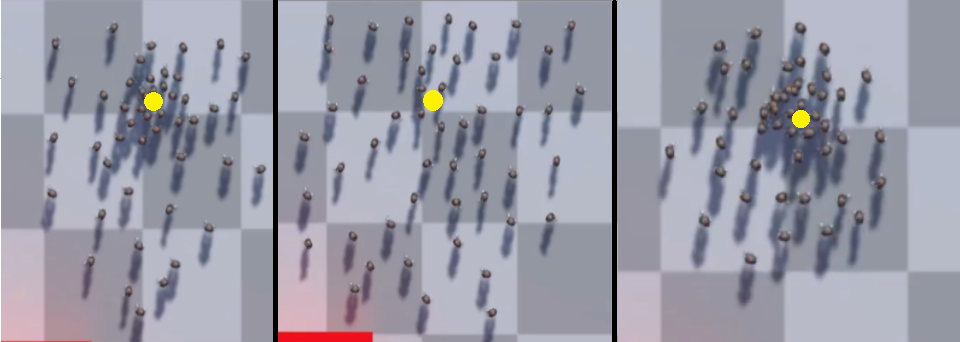}
  \caption{\red{Three applications of BioCrowds. The yellow dots represent the agents' goals. Left: BioCrowds with our proposed model of the Extraversion personality trait, using two distinct agent profiles: agents with $E = 1$ (closer to the goal and to each other); and agents with $E = 0.8$ (further from the goal and each other). Center: BioCrowds with Normal Life, as proposed by Rockenbach et al.~\cite{Rockenbach2018}. Right: the original BioCrowds model, as proposed by Bicho et al.~\cite{de2012simulating}.} 
  }
    \label{finalFrames}
\end{figure}

\red{Fig.~\ref{finalFrames}}
illustrates three situations using BioCrowds with 50 agents positioned around a goal. On the left, we have our extended model of BioCrowds with two different levels of Extraversion. We use 25 agents having 0.8 as Extraversion values and 25 agents with 1.0. In the center, we have the implementation of Normal Life model, according to Rockenbach et al.~\cite{Rockenbach2018}. It is easy to remark that agents are well distributed in the space trying to maximize their comfort. Finally, on the right, we have the original BioCrowds model.
It is easy to perceive how the personality changes the model results. On the left of 
\red{Fig.~\ref{finalFrames}},
one can perceive the distribution of agents, where the ones with higher values of Extraversion are close to each other, and also close to the goal, because they were not disturbed by the presence of other agents, so they went directly to the goal. Still, in the image on the left, we can see the agents with lower Extraversion far from the goal and far from each other, as well. It is important to notice that lower values of E are possible. However, agents, in those cases, can behave going far from the goal, and then subjects can easily perceive the difference. That is why we use values where agents still go towards the goal.   

\begin{figure*}[t!]
  \centering
  \subfigure[fig:ourModelHeteroA][Simulation frame 150.]{\includegraphics[width=0.31\textwidth,frame=1.0pt]{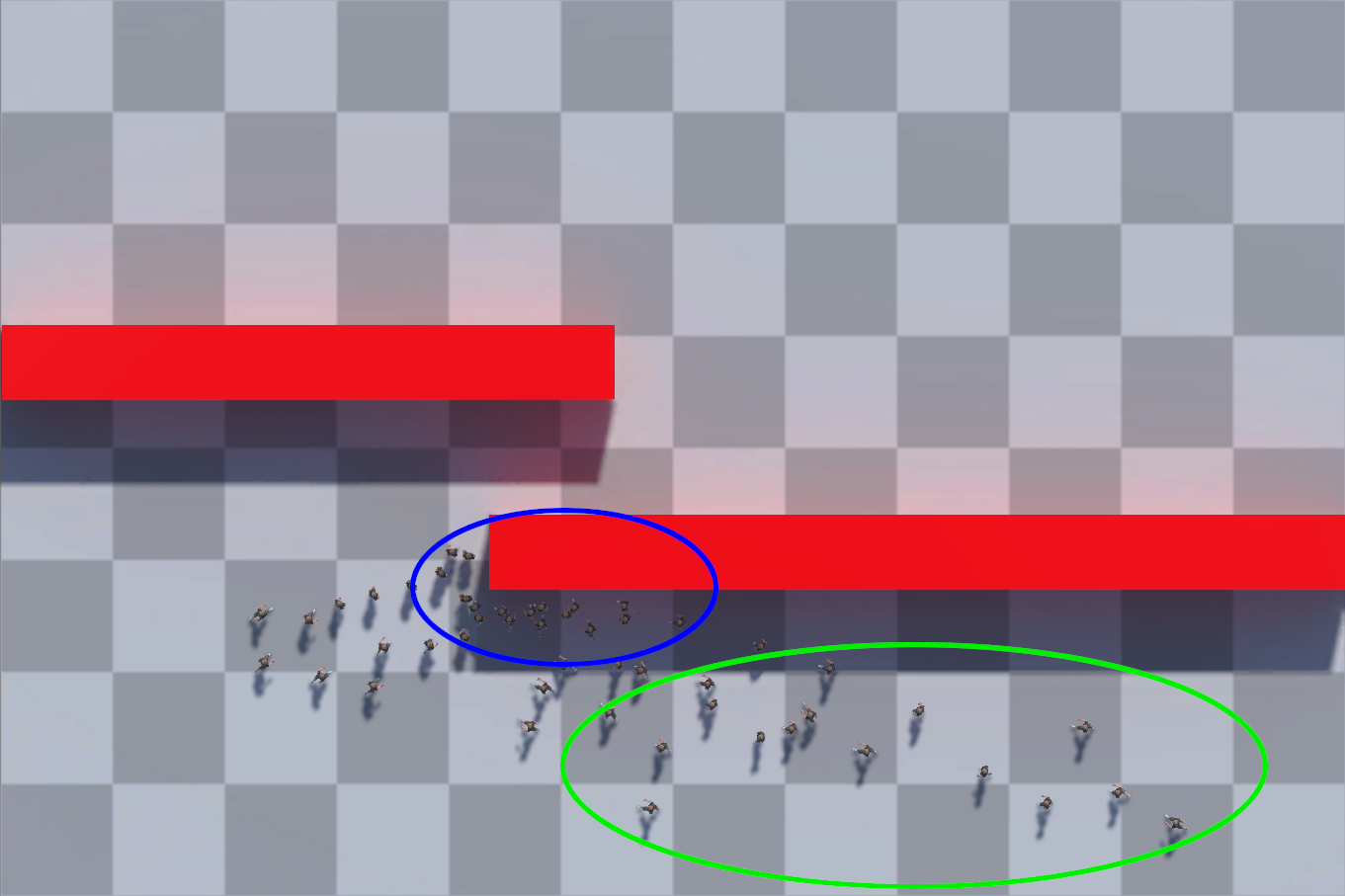}}
  \subfigure[fig:ourModelHeteroB][Simulation frame 450.]{\includegraphics[width=0.31\textwidth,frame=1.0pt]{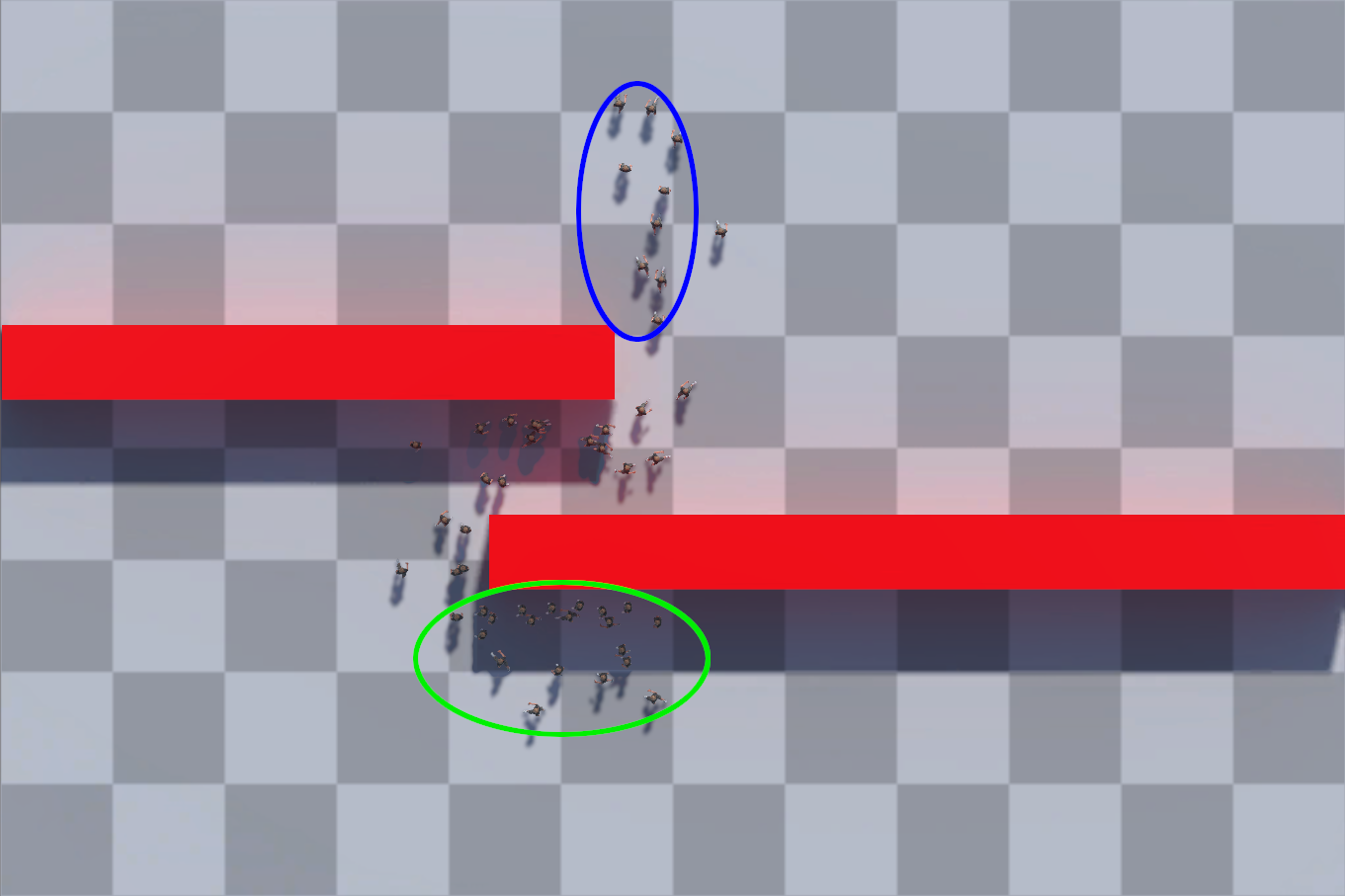} }
  \subfigure[fig:ourModelHeteroC][Simulation frame 1500.]{\includegraphics[width=0.31\textwidth,frame=1.0pt]{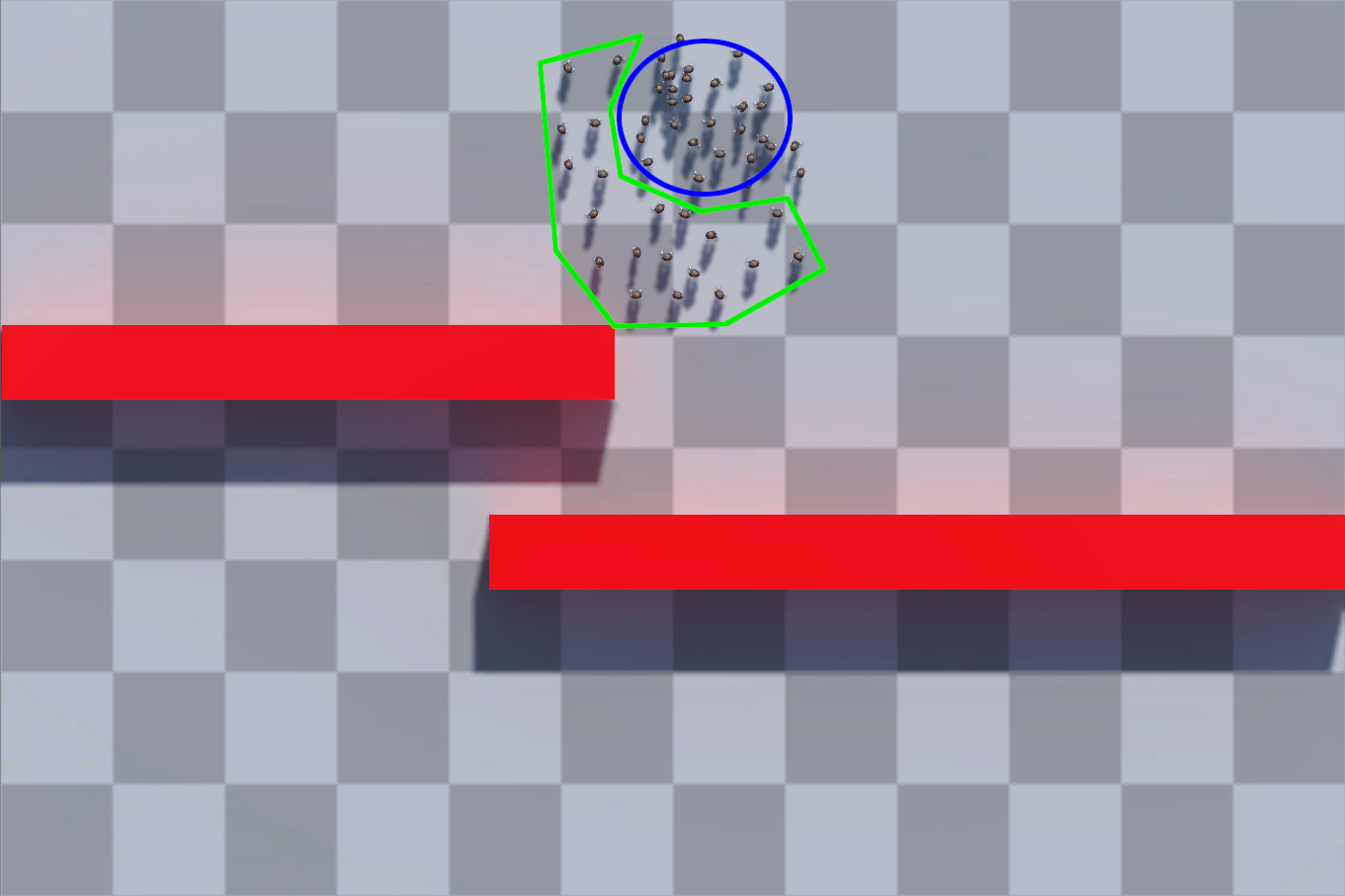}}
  \caption{\red{Evolution of a simulation utilizing our proposed model of the Extraversion (E) personality trait. Two agent profiles are presented: 25 agents with $E = 1$ (blue); and 25 agents with $E = 0.8$ (green). The frames 150 (a), 450 (b) and 1500 (c) are presented.
  }}
    \label{runFramesHeteronegeous}
\end{figure*}

\begin{figure*}[t!]
  \centering
  \subfigure[fig:ourModelHomo08A][Simulation frame 150.]{\includegraphics[width=0.31\textwidth,frame=1.0pt]{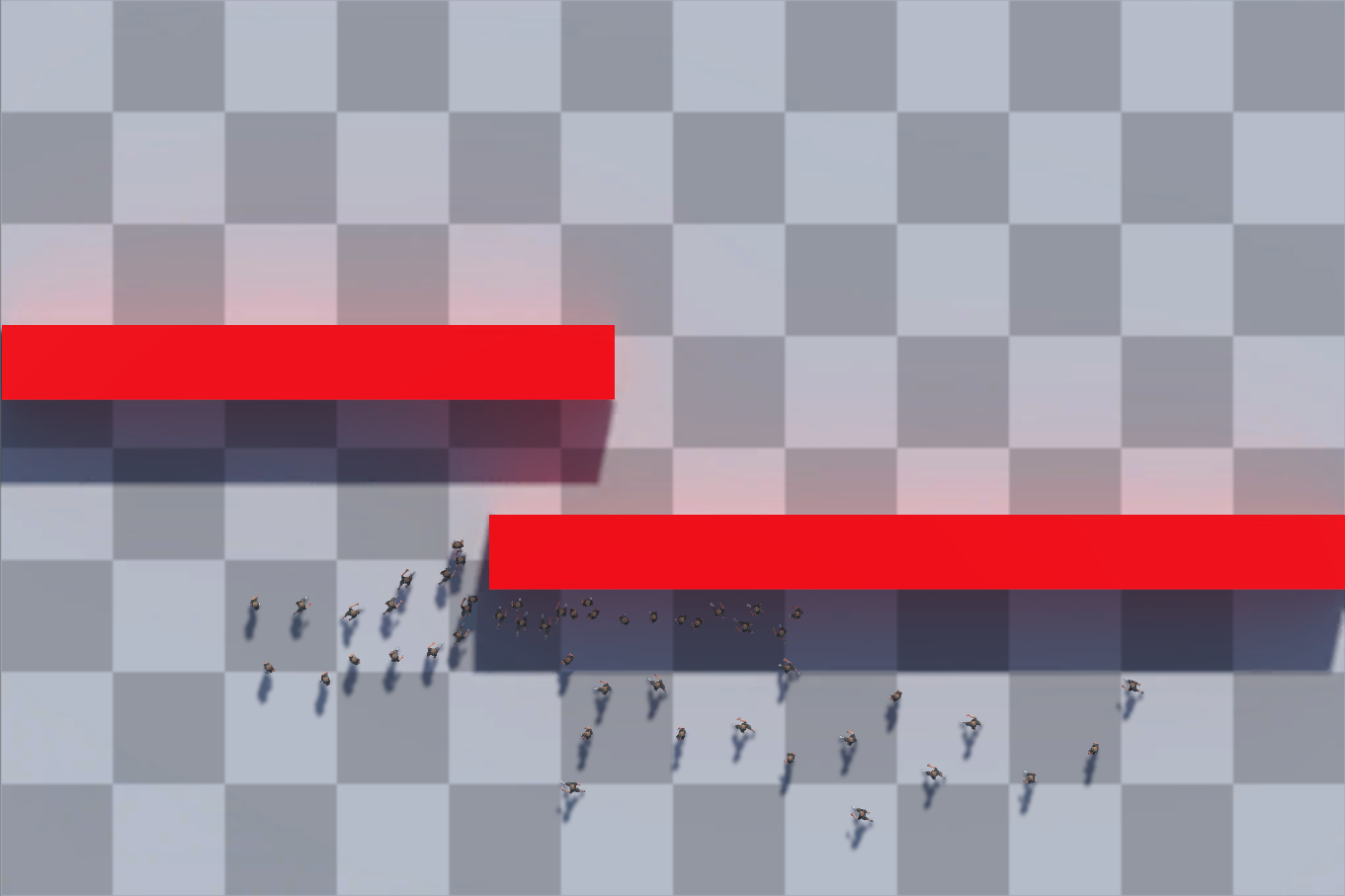}}
  \subfigure[fig:ourModelHomo08B][Simulation frame 450.]{\includegraphics[width=0.31\textwidth,frame=1.0pt]{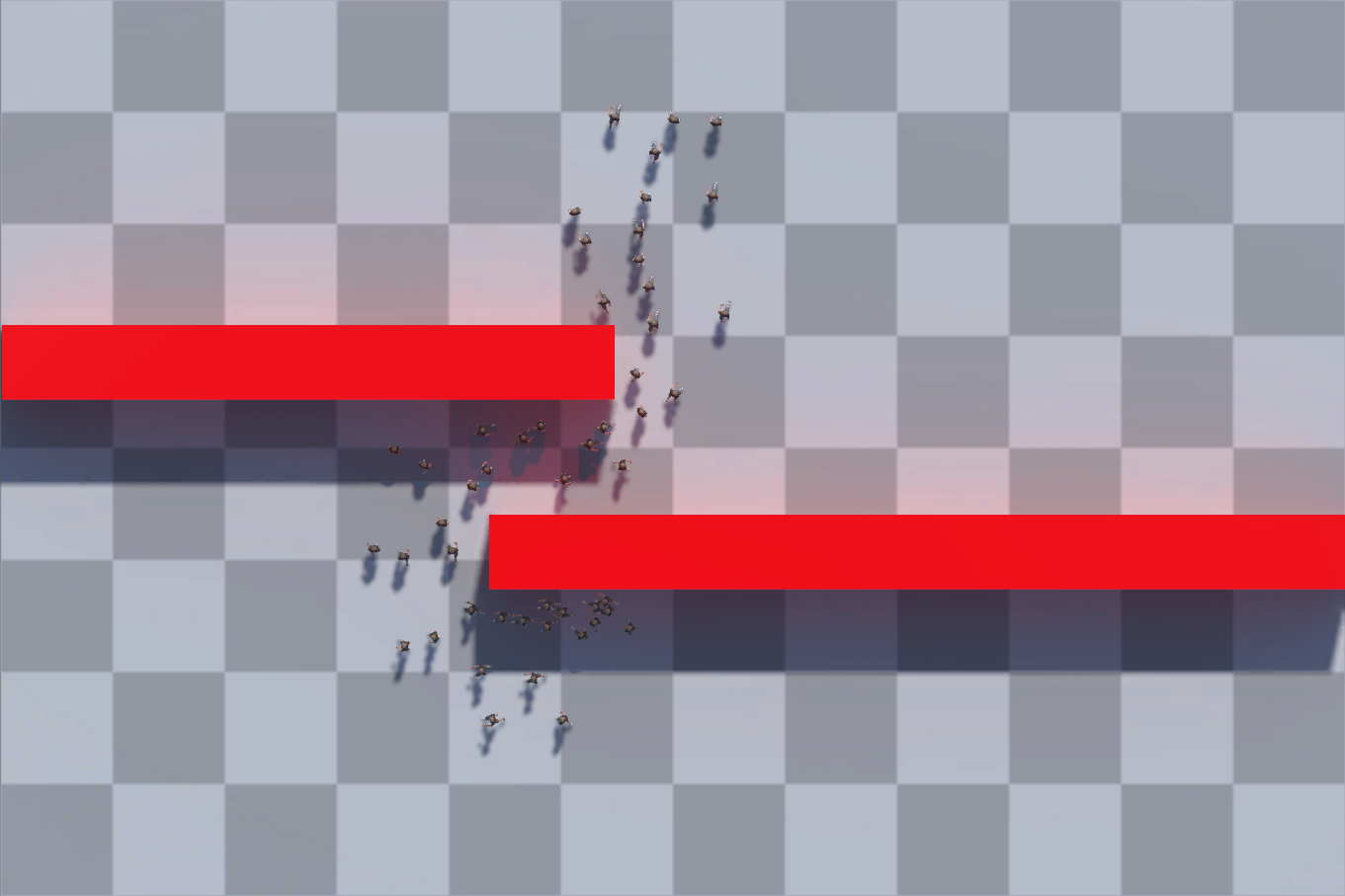}}
  \subfigure[fig:ourModelHomo08C][Simulation frame 1500.]{\includegraphics[width=0.31\textwidth,frame=1.0pt]{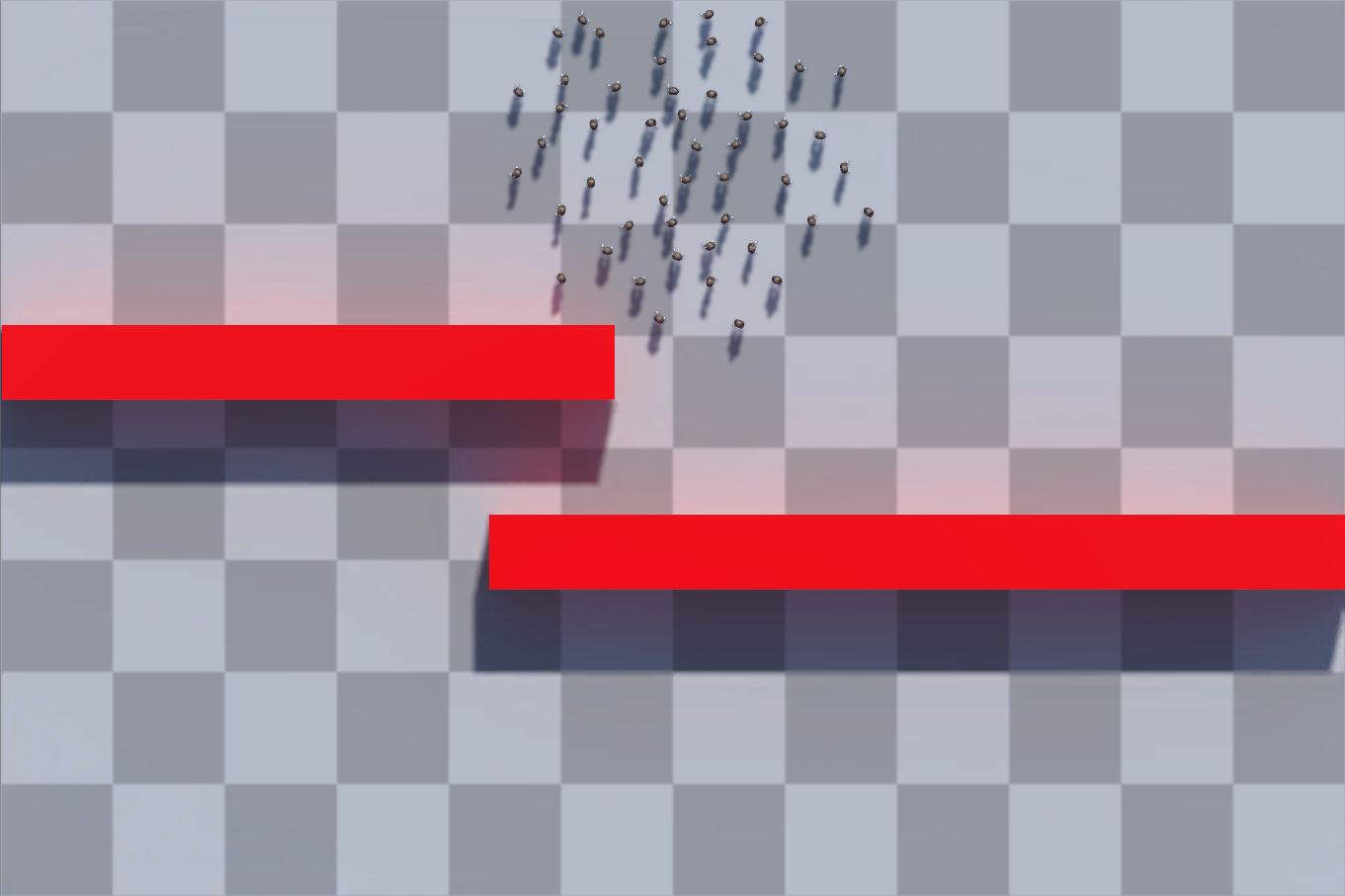}}
  \caption{\red{Evolution of a simulation utilizing our proposed model of the Extraversion (E) personality trait. All 50 agent present the value of $E = 0.8$. The frames 150 (a), 450 (b) and 1500 (c) are presented.
  }}
    \label{runFramesHomogeneous08}
\end{figure*}

\begin{figure*}[t!]
  \centering
  \subfigure[fig:normalLifeA][Simulation frame 150.]{\includegraphics[width=0.31\textwidth,frame=1.0pt]{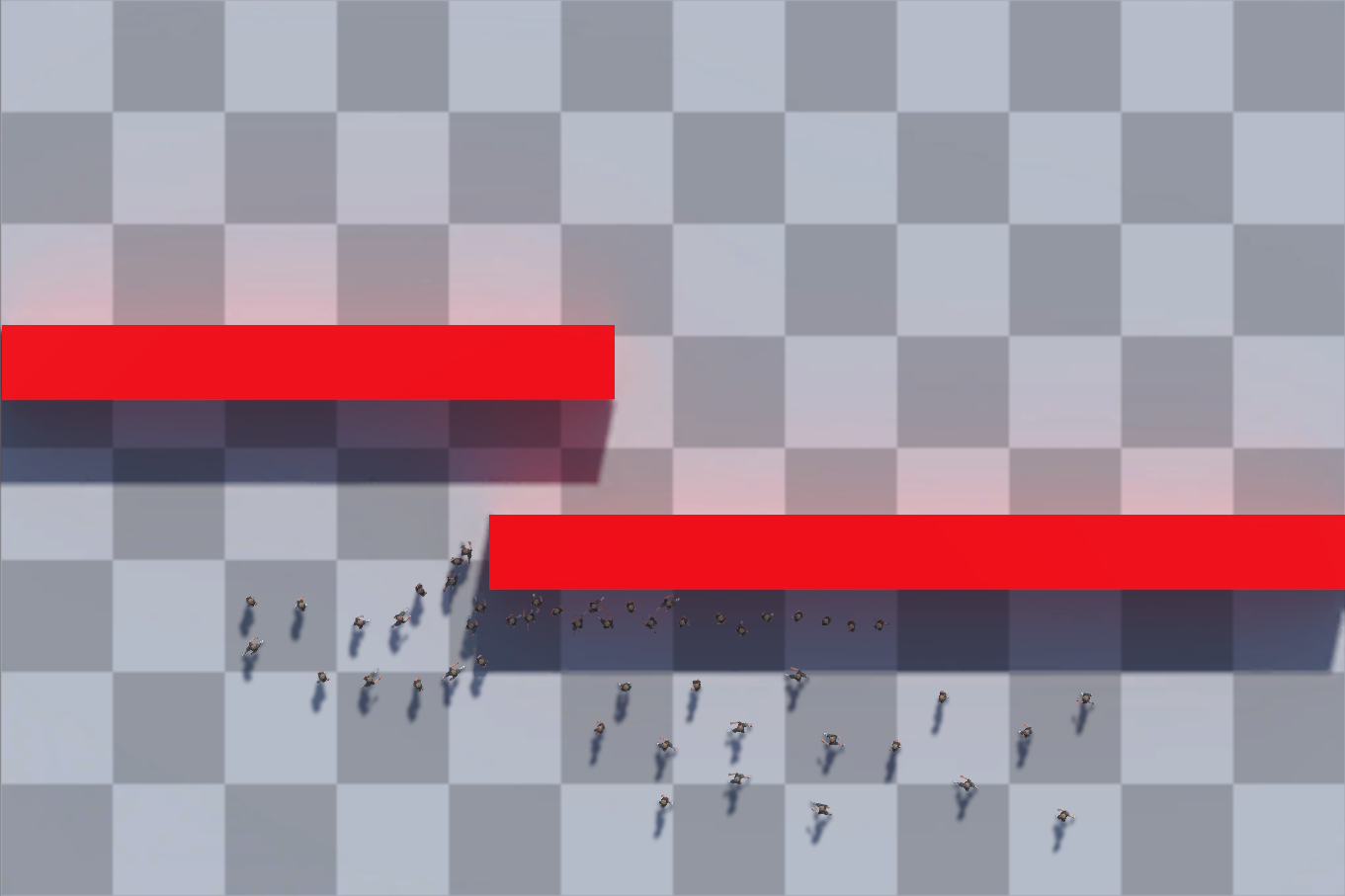}}
  \subfigure[fig:normalLifeB][Simulation frame 450.]{\includegraphics[width=0.31\textwidth,frame=1.0pt]{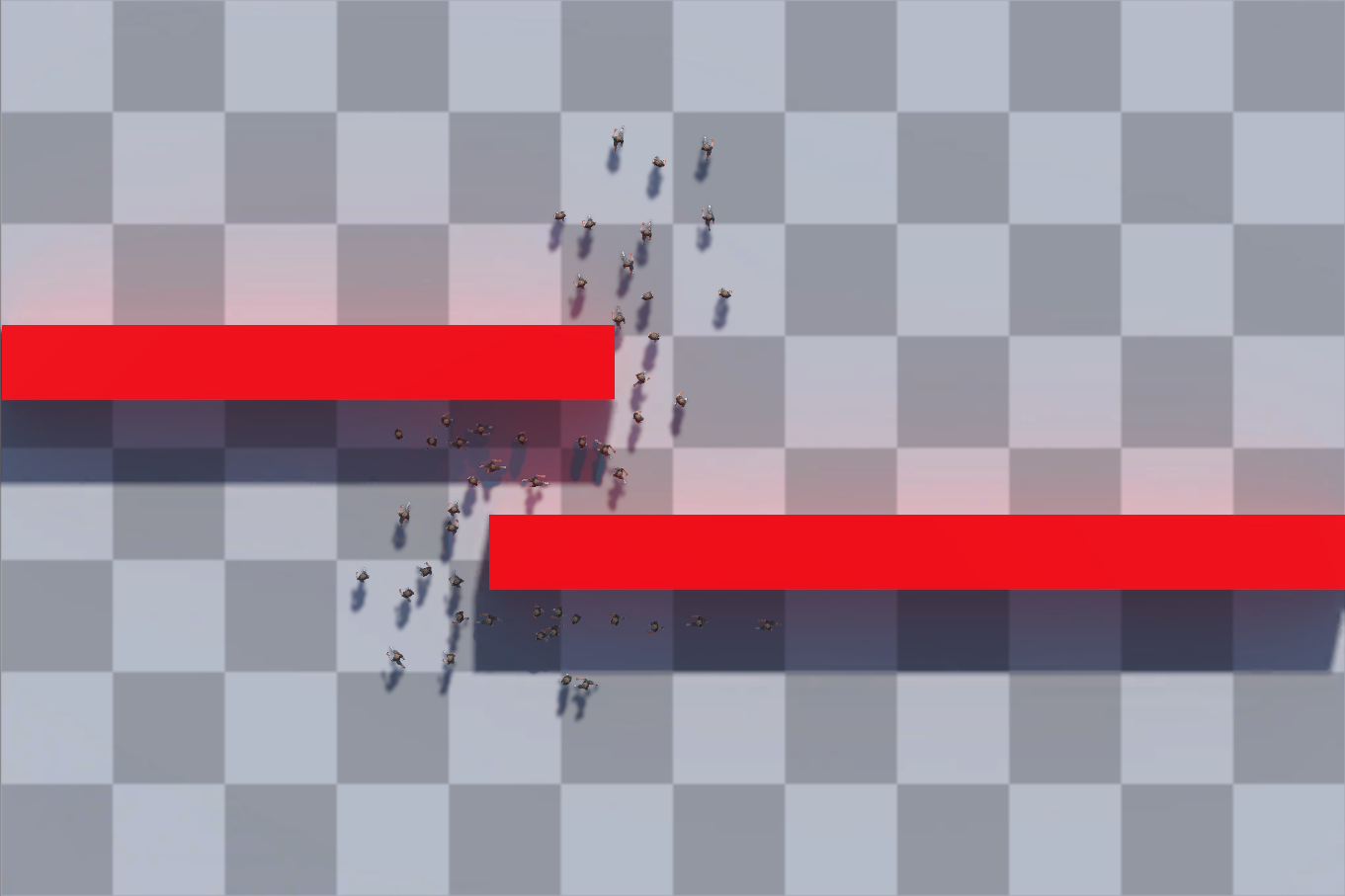}}
  \subfigure[fig:normalLifeC][Simulation frame 1500.]{\includegraphics[width=0.31\textwidth,frame=1.0pt]{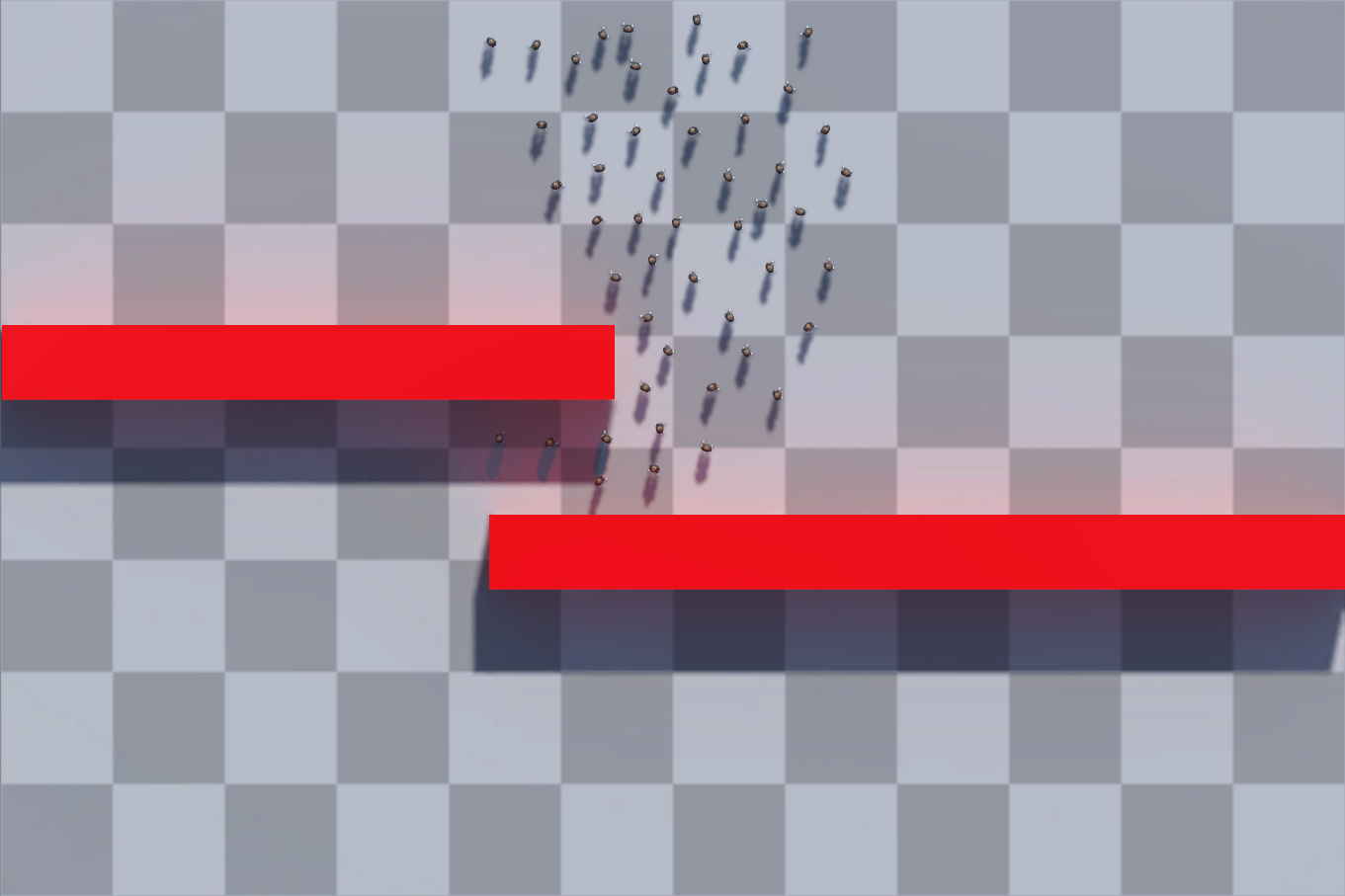}}
  \caption{\red{Evolution of a simulation utilizing the Normal Life extension model, as proposed by Rockenbach et al.~\cite{Rockenbach2018}. The frames 150 (a), 450 (b) and 1500 (c) are presented.
  }}
    \label{runFramesNormalLife}
\end{figure*}

\begin{figure*}[t!]
  \centering
  \subfigure[fig:bioCrowdsA][Simulation frame 150.]{\includegraphics[width=0.31\textwidth,frame=1.0pt]{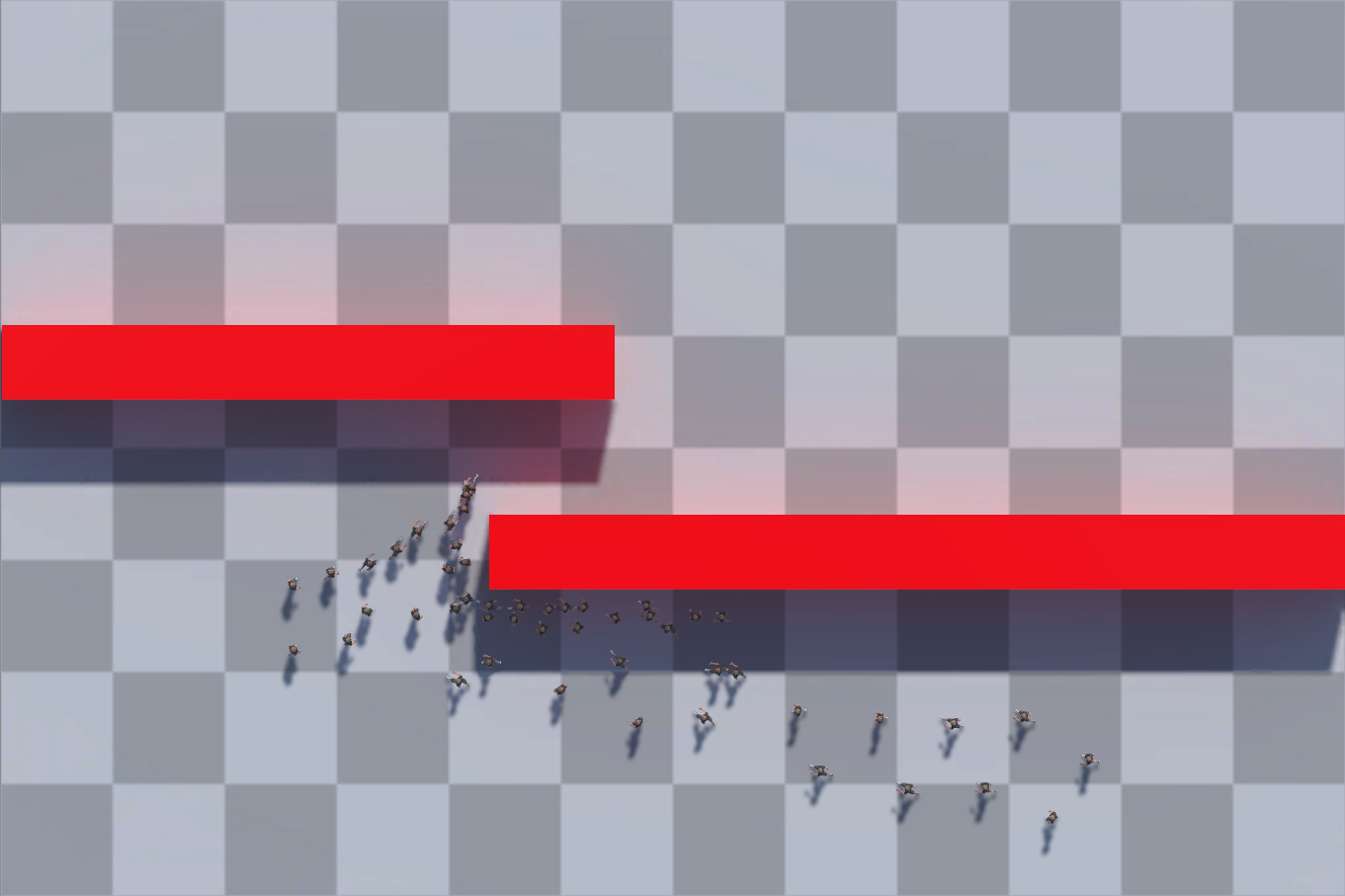}}
  \subfigure[fig:bioCrowdsB][Simulation frame 450.]{\includegraphics[width=0.31\textwidth,frame=1.0pt]{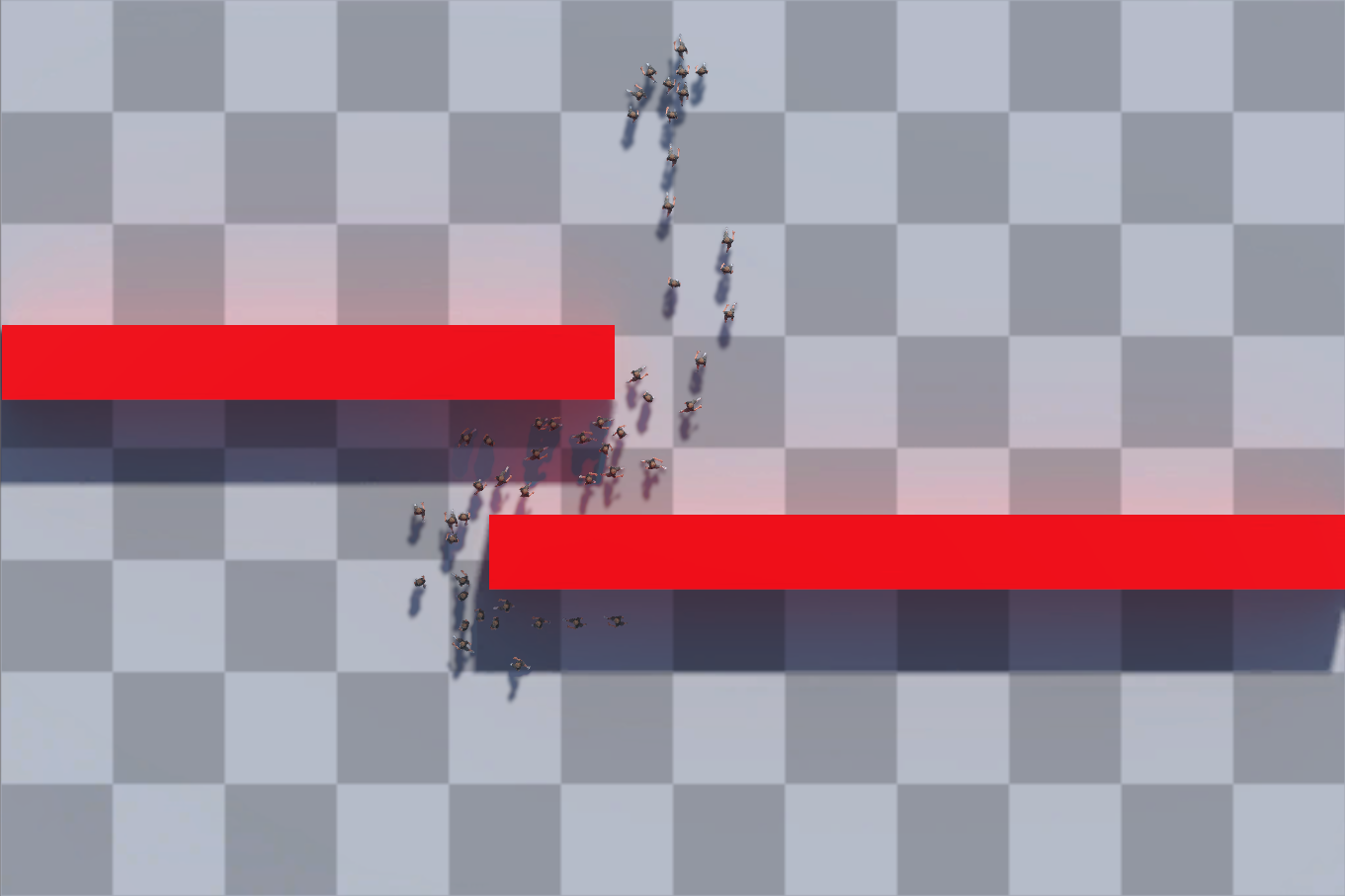}}
  \subfigure[fig:bioCrowdsC][Simulation frame 1500.]{\includegraphics[width=0.31\textwidth,frame=1.0pt]{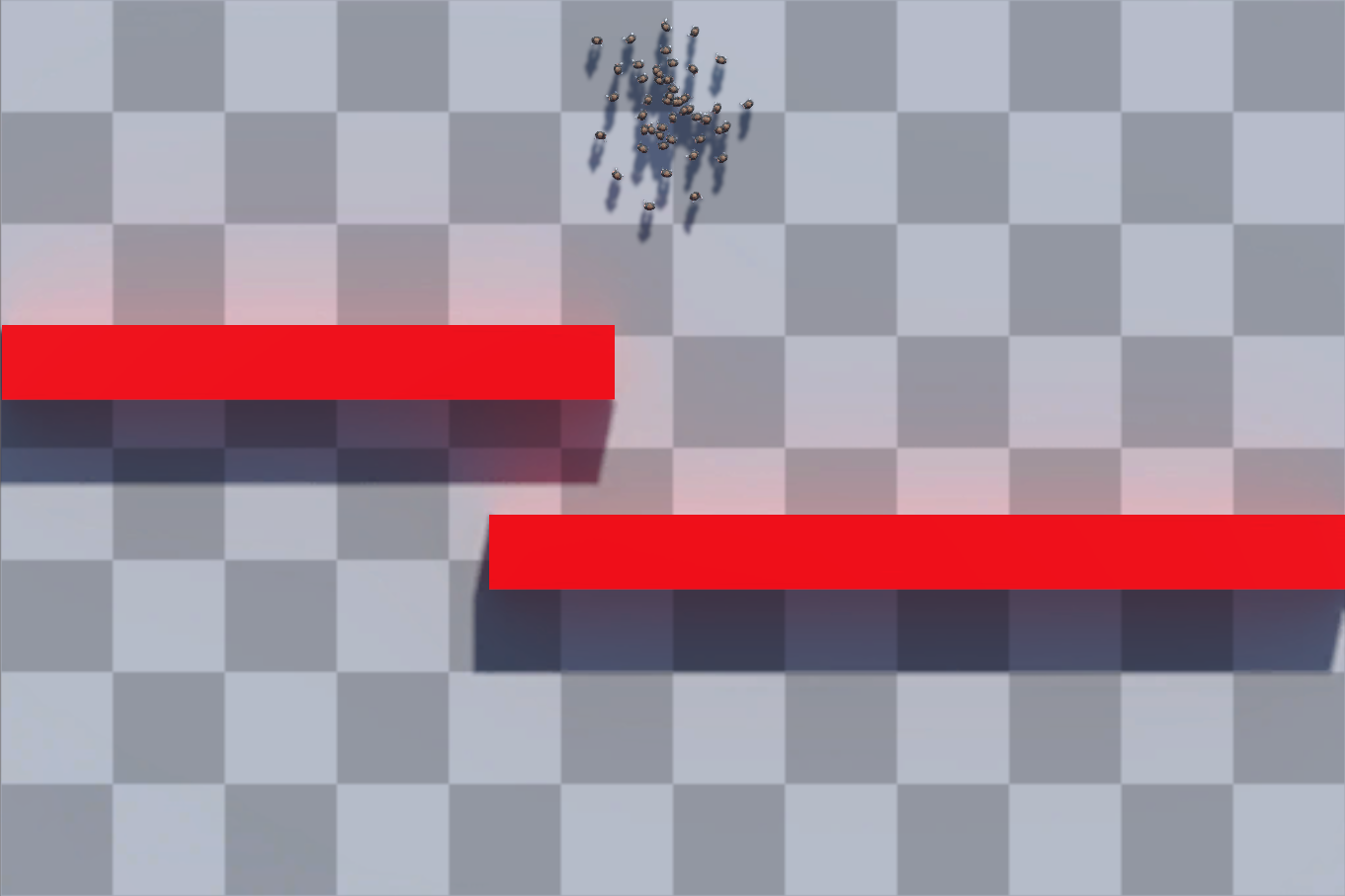}}
  \caption{\red{Evolution of a simulation utilizing the original BioCrowds model, as proposed by Bicho et al.~\cite{de2012simulating}. The frames 150 (a), 450 (b) and 1500 (c) are presented.
  }}
    \label{runFramesBioCrowds}
\end{figure*}

\red{Fig.~\ref{runFramesHeteronegeous} through~\ref{runFramesBioCrowds} present the evolution of four simulations, containing 50 agents each, using different  methods. Fig.~\ref{runFramesHeteronegeous} presents a simulation of our proposed model of Extraversion, using two distinct agent profiles: agents with $E = 1$ (highlighted in blue); and agents with $E = 0.8$ (highlighted in green). Fig.~\ref{runFramesHomogeneous08} presents a simulation of our proposed model of Extraversion with all agents having $E = 0.8$. Fig.~\ref{runFramesNormalLife} presents a simulation utilizing the Normal Life extension model, as proposed by Rockenbach et al.~\cite{Rockenbach2018}. Finally, Fig.~\ref{runFramesBioCrowds} presents a simulation utilizing the original BioCrowds model, as proposed by Bicho et al.~\cite{de2012simulating}.}

\red{In Fig.~\ref{runFramesHeteronegeous}(c), we can see that the agents with higher $E$ occupy less space, and tend to cluster together, whilst agents with lower $E$ occupy more space and keep a certain distance from each other. In Fig.~\ref{runFramesHomogeneous08}(c), we can observe agents with a lower value of Extraversion, where they tend to further themselves when disturbed by the presence of others, while still aiming for the goal. Similar behavior is perceived in Fig.~\ref{runFramesNormalLife}(c), with agents being more distributed in order to maximize their comfort.
Finally, Fig.~\ref{runFramesBioCrowds}(c) presents agents that do not take comfort and Extraversion into account, allowing them to be closer to one another and cluster around the goal.}





\section{Results}
\label{sec:results}
This section presents the obtained results when evaluating the people perception. 

\subsection{Research Methods}
\label{sec:ResearchMethods}
We developed a survey in Google Forms to understand how people perceive crowds personalities in our experiments. 
The survey was answered by 31 people, where 22.6\% are women and 74.4\% are men.
Other demographic attributes are following specified. Regarding the educational level, 58.06\% of the population have completed high school, and 41.96\% have higher education.
With respect to subjects' age, the average is $21.645$, therefore, people below and above average are respectively 80.6\% and 19.4\%. Subject with familiarity with CG is 22.6\% of the population, while 77.4\% declare themselves as non-familiar with CG. Initially, we informed people that they were free to give up responding in case of tiredness, boredom, or dizziness. In addition, we asked people if they agreed to participate into the survey. The experiments were organized in three scenarios: 
\\

\subsubsection{Scenario 1 - The user only observes the crowd}
In this scenario, the user observes the environment and the movement of the agents, using a spectator camera that does not affect the agents' behaviors. 

\subsubsection{Scenario 2 - The avatar is one agent in the original BioCrowds}
In this scenario, the user can 
interact with the agents whilst being able to occupy 
space while walking. In this Scenario, the agents take the user's presence into account and treat her/him as a BioCrowds agent.

\subsubsection{Scenario 3 - The avatar is one agent in BioCrowds Normal Life}
In this scenario, the user also can interact with the agents, as in Scenario 2, however, the agent wants to be comfortable in the space, as applied in the Normal Life model.
In this scenario, the agents consider the user's presence treating her/him as a Normal Life agent.  

The main difference between the avatar in Scenarios 2 and 3 is that in Scenario 2, agents or the avatar always replicate the main rule of BioCrowds, i.e., markers on the floor are attributed to the closest agent (weighted motion vectors use Equations~\ref{eq:biocrowds_motion} and~\ref{eq:biocrowds_marker_weight}. In Scenario 3, the markers are attributed to the agents (and the avatar) depending on their Extraversion values, as described in Equation~\ref{eq:extraversion}.
\\
\\
The scenarios were developed in the Unity3D engine and presented in a WebGL application. So, people accessed the scenarios through a GitHub link
distributed by Google Forms. We informed people that they could move freely between the survey link and the application link. After each scenario, the user had to answer two questions that reflect her/his perception throughout the simulation:


\begin{itemize}
    \item A) "Did you notice interactions between the agents?"
    \item B) "Did you perceive different emotions or personalities in the agents?"
    \\
\end{itemize}

Both questions were answered using 5-Likert Scales ("Did not notice at all" to "Noticed completely"), as shown in 
\red{Fig.~\ref{answerA} and~\ref{answerB}}.
Question A was asked to evaluate $H0_1$, question B to evaluate $H0_2$. With respect to $H0_3$, we measured the relationships between the results of questions A and B. The next section presents our findings with respect to the applied experiments and surveys.

\subsection{Research Results}
\label{sec:ResearchResults}

\begin{figure}[t!]
  \centering
  \includegraphics[scale=1.0, width=\linewidth]{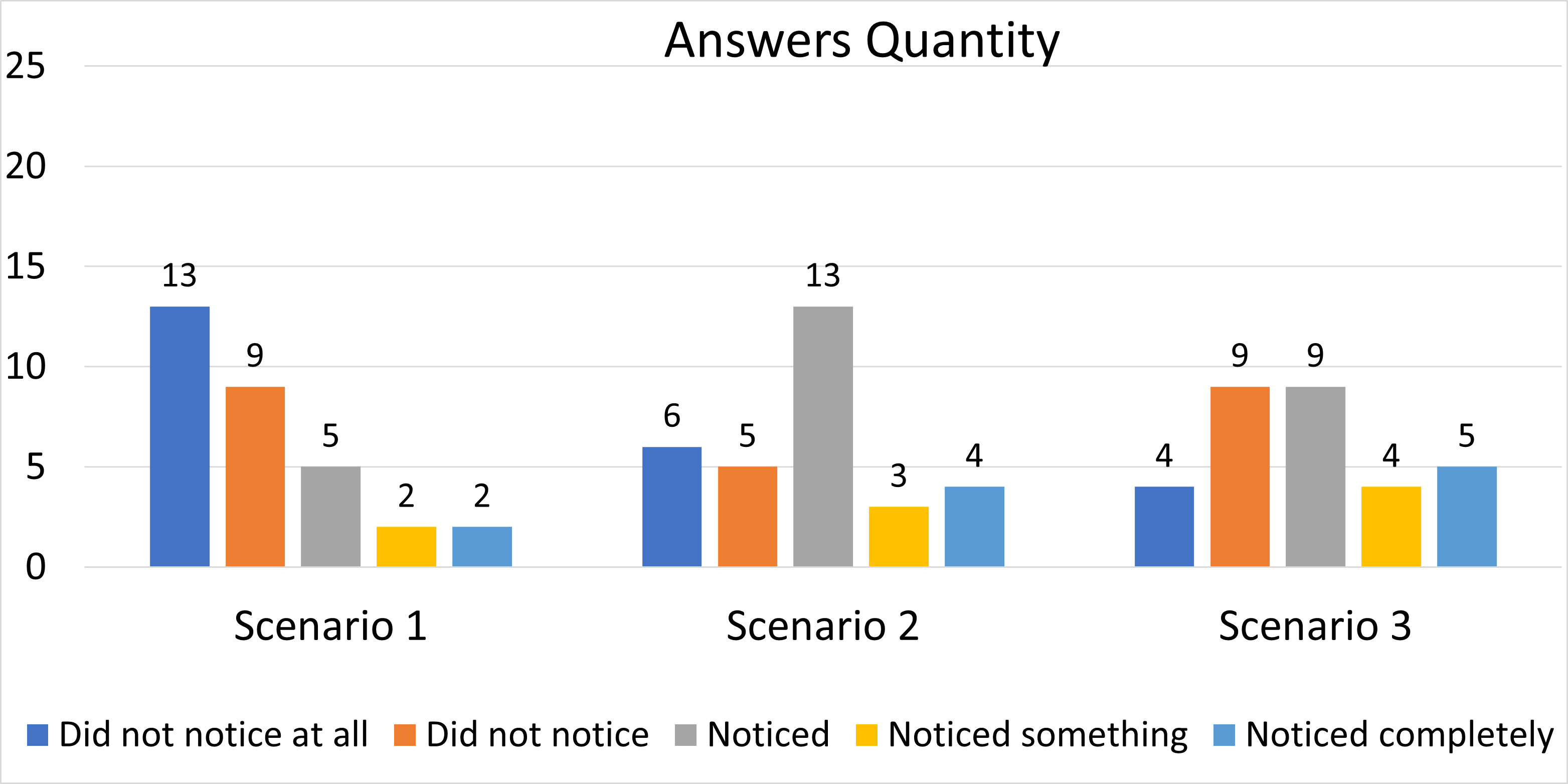}
  \caption{Answers collected from the form question ("Did you notice interactions between the agents?"), regarding Scenarios 1, 2 and 3.}
    \label{answerA}
\end{figure}

\begin{figure}[t!]
  \centering
  \includegraphics[scale=1.0, width=\linewidth]{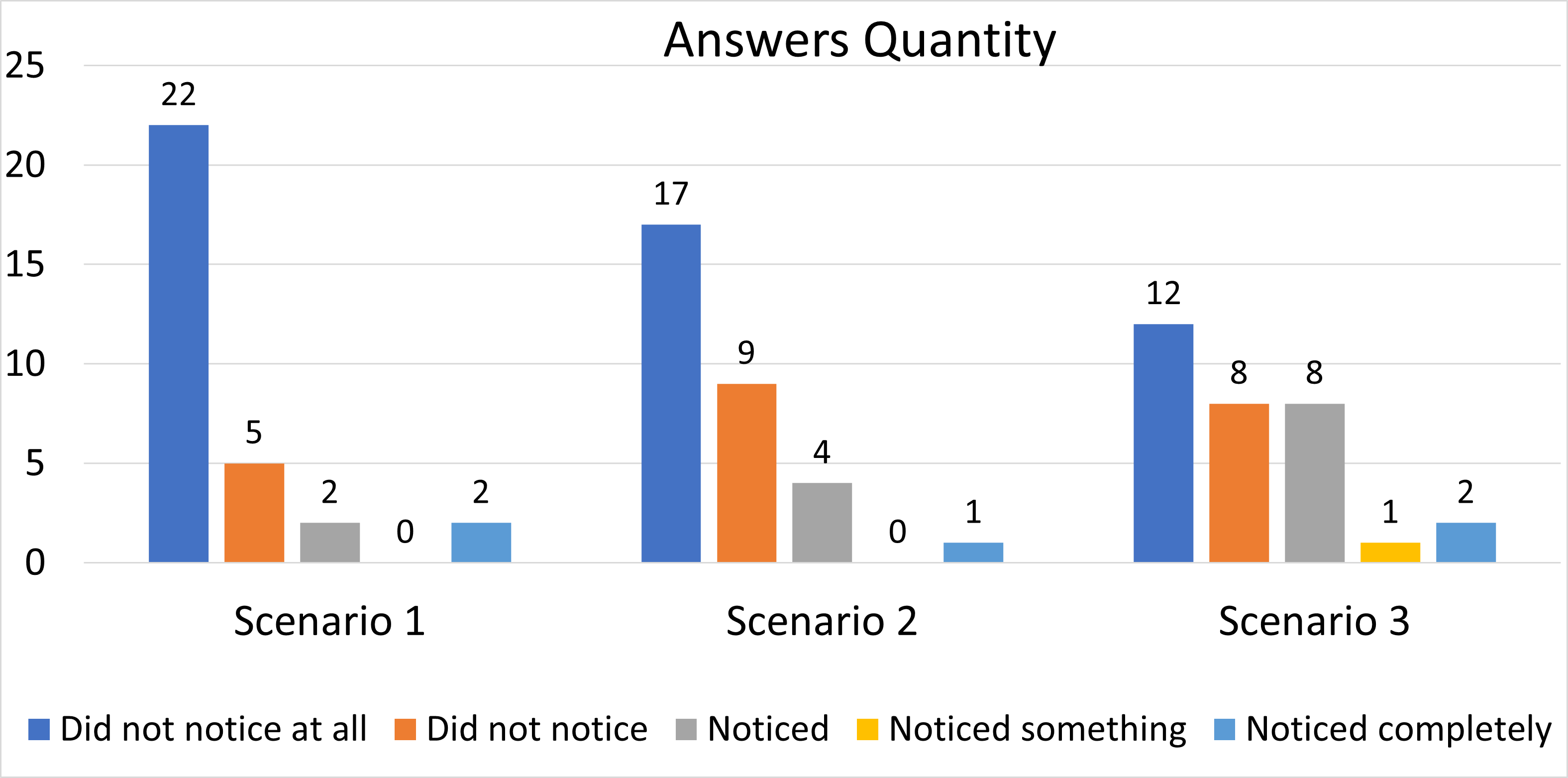}
  \caption{Answers collected from the form question ("Did you perceive different emotions or personalities between the agents?"), regarding Scenarios 1, 2 and 3.}
    \label{answerB}
\end{figure}

Based on our results, we found that the perception of the \red{extraversion personality trait} of virtual agents and the interaction between them depends on the user's form of interaction, as further discussed in this section.
In Scenario 1, where the user could only observe the agents, few users perceived interactions between agents or agent's emotions and personalities. On the other hand, in Scenarios 2 and 3, in which users could interact with agents with a virtual physical body, the users perceived agents interacting among themselves, as shown in 
\red{Fig.~\ref{answerA}},
but had difficulties identifying emotions and personalities, as shown in 
\red{Fig.~\ref{answerB}}.


In addition, in order to answer the hypotheses presented in Section~\ref{sec:introduction}, we performed statistical analysis using the \textit{Mann-Whitney} test of hypotheses (to evaluate $H0_1$ and $H0_2$) and \textit{Spearman} correlations (to evaluate $HO_3$) through the Scipy library in the Python language, using 95\% significance level. Both the Mann-Whitney test and the Spearman correlation were used because they are robust methods for unbalanced samples, such as the percentage of male participants was higher than the percentage of female participants obtained in our results. For these analyses, we scored the Likert Scales from 1 (Did not notice at all) to 5 (Noticed completely), and for the hypothesis tests, we used the averages of these scores (averages presented in Table~\ref{tbl:likertAvg}). We performed a general analysis and using demographic profiles. In the hypotheses tests, we compared the results as follows: \textit{i)} Relating to the applied scenarios, for example, the average of the Likert scores answers from question A in Scenario 1 vs. the average of the answers from question A in Scenario 2. So this was made using the three Scenarios (1, 2, 3) x the two questions (A and B) x demographics (gender, familiarity with CG, education level, age), resulting in 24 analyzed configurations; 
\textit{ii)} Relating to the demographic data, for example, the average of women's answers to question A x the average of men's answers to question A. In this case, we compared the overall averages, that is, answers to questions A and B taking into account all Scenarios (1, 2, 3), and the averages of the questions taking into account the scenarios separately. With respect to correlations, we measured the relationships between questions A's answers and B's answers.
\begin{table}[!htb]
    \centering
    \caption{Table of average of questions A and B (using Likert scales as scores) in all analysis (general, gender, education, age, familiarity with CG).}
    \label{tbl:likertAvg}
    
    \begin{adjustbox}{max width=0.98\linewidth}
    \begin{tabular}{|c|c|c|c|}
        \hline
       \multirow{2}{*}{\shortstack[c]{\textbf{Analysis}}} & \multirow{2}{*}{\shortstack[c]{\textbf{Scenario}}}  & \textbf{Question A} & \textbf{Question B} \\ 
       & & \textbf{(AVG)} & \textbf{(AVG)} \\
        
        \hline
        \textbf{General} & All (1,2,3) & $2.473$ & $2.043$ \\
        \hline
        
        \textbf{General} & 1 & $2.129$ & $2.032$ \\
        \hline
        
        \textbf{General} & 2 & $2.71$ & $1.903$ \\
        \hline
        
        \textbf{General} & 3 & $2.581$ & $2.194$ \\
        \hline
        \noalign{\smallskip}
        \hline
        
       \textbf{Analysis} & \multirow{2}{*}{\shortstack[c]{\textbf{Scenario}}}  & \textbf{Question A} & \textbf{Question B} \\ 
      \textbf{(Gender)} & & \textbf{(AVG)} & \textbf{(AVG)} \\
       
       \hline
        
        \textbf{Women} & All (1,2,3) & $2.476$ & $2.449$ \\
        \hline
        
        \textbf{Women} & 1 & $2.143$ & $2.286$ \\
        \hline
        
        \textbf{Women} & 2 & $2.714$ & $2.143$ \\
        \hline
        
        \textbf{Women} & 3 & $2.571$ & $2.286$ \\
        \hline

        \textbf{Men} & All (1,2,3) & $2.449$ & $1.942$ \\
        \hline
        
        \textbf{Men} & 1 & $2.087$ & $1.913$ \\
        \hline
        
        \textbf{Men} & 2 & $2.696$ & $1.783$ \\
        \hline
        
        \textbf{Men} & 3 & $2.565$ & $2.13$ \\
        \hline
        \noalign{\smallskip}
        \hline
        
        \textbf{Analysis} & \multirow{2}{*}{\shortstack[c]{\textbf{Scenario}}}  & \textbf{Question A} & \textbf{Question B} \\ 
       \textbf{(Education)} & & \textbf{(AVG)} & \textbf{(AVG)} \\
       
       \hline
        
        \textbf{Complete} & \multirow{2}{*}{\shortstack[c]{All (1,2,3)}} & \multirow{2}{*}{\shortstack[c]{$2.389$}} & \multirow{2}{*}{\shortstack[c]{$2.185$}} \\
        \textbf{High School} & & & \\
        \hline
        
        \textbf{Complete} & \multirow{2}{*}{\shortstack[c]{1}} & \multirow{2}{*}{\shortstack[c]{$2.167$}} & \multirow{2}{*}{\shortstack[c]{$2.111$}} \\
        \textbf{High School} & & & \\
        \hline
        
        \textbf{Complete} & \multirow{2}{*}{\shortstack[c]{2}} & \multirow{2}{*}{\shortstack[c]{$2.5$}} & \multirow{2}{*}{\shortstack[c]{$2.111$}} \\
        \textbf{High School} & & & \\
        \hline
        
        \textbf{Complete} & \multirow{2}{*}{\shortstack[c]{3}} & \multirow{2}{*}{\shortstack[c]{$2.5$}} & \multirow{2}{*}{\shortstack[c]{$2.333$}} \\
        \textbf{High School} & & & \\
        \hline

        \textbf{Higher} & \multirow{2}{*}{\shortstack[c]{All (1,2,3)}} & \multirow{2}{*}{\shortstack[c]{$2.59$}} & \multirow{2}{*}{\shortstack[c]{$1.846$}} \\
        \textbf{Education} & & & \\
        \hline
        
        \textbf{Higher} & \multirow{2}{*}{\shortstack[c]{1}} & \multirow{2}{*}{\shortstack[c]{$2.077$}} & \multirow{2}{*}{\shortstack[c]{$1.923$}} \\
        \textbf{Education} & & & \\
        \hline
        
        \textbf{Higher} & \multirow{2}{*}{\shortstack[c]{2}} & \multirow{2}{*}{\shortstack[c]{$3.0$}} & \multirow{2}{*}{\shortstack[c]{$1.165$}} \\
        \textbf{Education} & & & \\
        \hline
        
        \textbf{Higher} & \multirow{2}{*}{\shortstack[c]{3}} & \multirow{2}{*}{\shortstack[c]{$2.692$}} & \multirow{2}{*}{\shortstack[c]{$2.0$}} \\
        \textbf{Education} & & & \\
        \hline
        \noalign{\smallskip}
        \hline
        
        \textbf{Analysis} & \multirow{2}{*}{\shortstack[c]{\textbf{Scenario}}}  & \textbf{Question A} & \textbf{Question B} \\ 
        \textbf{(Age)} & & \textbf{(AVG)} & \textbf{(AVG)} \\
       
       \hline
        
        \textbf{$\bf< 21.645$} & All (1,2,3) & $2.373$ & $2.08$ \\
        \hline
        
        \textbf{$\bf< 21.645$} & 1 & $2.12$ & $2.04$ \\
        \hline
        
        \textbf{$\bf< 21.645$} & 2 & $2.6$ & $2.0$ \\
        \hline
        
        \textbf{$\bf< 21.645$} & 3 & $2.4$ & $2.2$ \\
        \hline
        
        \textbf{$\bf> 21.645$} & All (1,2,3) & $2.889$ & $1.889$ \\
        \hline
        
        \textbf{$\bf> 21.645$} & 1 & $2.167$ & $2.0$ \\
        \hline
        
        \textbf{$\bf> 21.645$} & 2 & $3.167$ & $1.5$ \\
        \hline
        
        \textbf{$\bf> 21.645$} & 3 & $3.333$ & $2.167$ \\
        \hline
        \noalign{\smallskip}
        \hline
        
       \textbf{Analysis} & \multirow{2}{*}{\shortstack[c]{\textbf{Scenario}}}  & \textbf{Question A} & \textbf{Question B} \\ 
       \textbf{(CG Familiarity)} & & \textbf{(AVG)} & \textbf{(AVG)} \\
       
       \hline
        
        \textbf{With Familiarity} & All (1,2,3) & $2.762$ & $2.095$ \\
        \hline
        
        \textbf{With Familiarity} & 1 & $2.286$ & $2.286$ \\
        \hline
        
        \textbf{With Familiarity} & 2 & $3.0$ & $2.0$ \\
        \hline
        
        \textbf{With Familiarity} & 3 & $3.0$ & $2.571$ \\
        \hline
        
        \textbf{Did not know or} & \multirow{2}{*}{\shortstack[c]{All (1,2,3)}} & \multirow{2}{*}{\shortstack[c]{$2.389$}} & \multirow{2}{*}{\shortstack[c]{$1.972$}} \\
        \textbf{not familiar} & & & \\
        \hline
        
        \textbf{Did not know or} & \multirow{2}{*}{\shortstack[c]{1}} & \multirow{2}{*}{\shortstack[c]{$2.083$}} & \multirow{2}{*}{\shortstack[c]{$1.958$}} \\
        \textbf{not familiar} & & &\\
        \hline
        
        \textbf{Did not know or} & \multirow{2}{*}{\shortstack[c]{2}} & \multirow{2}{*}{\shortstack[c]{$2.625$}} & \multirow{2}{*}{\shortstack[c]{$1.875$}} \\
        \textbf{not familiar} & & &  \\
        \hline
        
        \textbf{Did not know or} & \multirow{2}{*}{\shortstack[c]{3}} & \multirow{2}{*}{\shortstack[c]{$2.458$}} & \multirow{2}{*}{\shortstack[c]{$2.083$}} \\
        \textbf{not familiar} & & & \\
        \hline

    \end{tabular}
    \end{adjustbox}
\end{table}
Regarding the general analysis, in the first four lines of Table~\ref{tbl:likertAvg}, (without separating into demographic data), we only found significant results when comparing the averages of question A ($H0_1$) between Scenarios 1 and 2 (\textit{p}-value $.01$). Therefore, \textbf{we can say that people perceived more interactions between agents in Scenario 2 than in Scenario 1.} Regarding Spearman's correlations, we found two significant \textit{p}-values in the general ($.018$ in all Scenarios) and Scenario 3 ($.03$) analysis between questions A and B ($H0_3$). However, the correlation values were low, being $.245$ in the general and $.39$ in Scenario 3. As in the general analysis, the correlation value was below $.3$, so \textbf{we can say that there was a weak correlation in Scenario 3 between the answers of A and B. Therefore, in Scenario 3, we can say that there was a weak tendency that the more people perceived interactions, the more they perceived that agents had different personalities and emotions (and vice versa).}

With respect to gender (we excluded a person from this analysis, as that person did not declare their gender), we did not find significant results when we evaluated the women's data (both in the hypothesis tests and in the correlations). In this case, we may not have found significant results because the number of female participants was very low compared to the number of male participants.
As in the general analysis, we only found a significant result when we compared the averages of question A ($H0_1$) between Scenarios 1 and 2 ($.018$). So, \textbf{we can say that men perceived more interactions in Scenario 2 than in Scenario 1.} The results of the correlations were also similar to the results of the general analysis, that is, significant \textit{p}-values in the correlations between questions A and B ($H0_3$) taking into account all scenarios together ($.036$), and taking into account only Scenario 3 ($.03$). The correlation values were, respectively, $.253$ and $.454$. Therefore, \textbf{looking only at the correlation between A and B's answers in Scenario 3, we can say that there was a weak trend that the more men perceived the interactions, the more they perceived that agents had different personalities and emotions (and vice versa).} We did not find significant results in comparisons between data from women vs. data from men.

With respect to educational level, for people with complete high school, we found significant \textit{p}-values ($.002$ and $.046$) in the weak correlations ($.409$ and $.475$) between questions A and B ($H0_3$) for the scenarios in general and Scenario 3 separately. With that, \textbf{we can say that in general and in Scenario 3, there was a weak tendency that the more people with complete high school perceived the interactions, the more they perceived that the agents had different personalities and emotions (and vice versa).} For people with higher education, we only found a significant result ($.01$) in the comparison between Scenarios 1 and 2 in question A ($H0_1$). Therefore, \textbf{we can say that people with higher education perceived more interactions in Scenario 2 than in 1.} Comparing the answers of people with complete high school Vs. people with higher education, we found significant results in the comparisons of question B ($H0_2$) related to the general analysis ($.037$) and the analysis of Scenario 2 ($.034$). With this, \textbf{in general (and in Scenario 2), we can say that people with complete high school perceived more different personalities and emotions in agents than people with higher education.}

Regarding age, for below-average people, we found a significant result ($.037$) in the comparison between Scenarios 1 and 2 in question A ($H0_1$). We also found a significant \textit{p}-value ($.016$) in the general correlation (all Scenarios) between questions A and B ($H0_3$), but the correlation value ($.278$) was too low. Therefore, \textbf{we can only say that people below the average age perceived more interactions between agents in Scenario 2 than in Scenario 1.} Regarding above-average people, we only found significant results ($.035$) when comparing Scenarios 1 and 2 in question B ($H0_2$). Therefore, \textbf{we can say that above the average age people perceived more different personalities and emotions in agents in Scenario 1 than in Scenario 2.} 
Comparing people below the average age vs. above, we only found significant results in question A ($H0_1$) when we analyzed in general ($.037$) and in Scenario 3 ($.036$). With that, \textbf{we can say that in general (and in Scenario 3), above the average age people perceived more interactions between agents than below.}

Regarding familiarity with CG, we only found a significant \textit{p}-value in the correlation between the answers to questions A and B ($H0_3$) in Scenario 3 ($.026$), and differently from the previous results, having a strong correlation of $.814$. Thus, \textbf{in Scenario 3, we can say that there was a strong tendency that the more people familiar with CG perceived interactions, the more they perceived that agents had different personalities or emotions (vice versa).} 

Regarding people who did not know or were not familiar with CG, we only found a significant result in the comparison between Scenarios 1 and 2 in question A ($.02$), i.e., $H0_1$. With that, \textbf{we can say that these people perceived more interactions in Scenario 2 than in Scenario 1.} In the comparison between the two groups (people with familiarity with CG vs. people who did not know or were not familiar with CG), we did not find significant results.

\section{Discussion}
\label{sec:discussion}

In this section, we report our discussions of the results presented in the previous section 
with respect to the three research hypotheses. Remembering, the hypotheses are: \textit{i)} $H0_1$  defining that people with only observational control of agents in the crowd (do not interfere with crowd dynamics) perceive interactions similarly to people with control of agents in the crowd (the user is considered a crowd agent); \textit{ii)} $H0_2$ defining that people with only observational control of crowd agents perceive different personalities and emotions similarly to people with control of crowd agents; \textit{iii)} $H0_3$ defining that the perception of interactions in crowds is not related to the perception of different personalities and emotions.

Regarding $H0_1$ (perception of interactions), people in general (also separately - men, higher education, people below the average age, and who did not know or were not familiar with CG), perceived more interactions in Scenario 2 than in Scenario 1. In the other comparisons between scenarios (1 vs. 3 and 2 vs. 3), we did not find significant results. However, if we look only at the averages of the general analysis in Table~\ref{tbl:likertAvg}, we can see that Scenario 2 had the highest average values of perception in question A. These results refute $H0_1$ and tell us that people perceived more interactions when they were part of the interactions (looking for their spaces) than when they just watched the agents interacting. In addition, we found an age effect, where people above average age perceived more interactions than people below average age.

Regarding $H0_2$ (perception of different personalities and emotions), we only found significant results when we separated people by age. People who were above the average age perceived more different personalities and emotions in Scenario 1 than in Scenario 2. This result is interesting because users only observed the agents in scenario 1, that is, the camera did not influence the interactions. However, if we look only at the averages (Table~\ref{tbl:likertAvg}), we can see that all question B average values of perception of different personalities and emotions
for Scenario 2 were the lowest compared to the other scenarios. Scenario 3 was the one with the highest perception values of different personalities and emotions. These results may indicate that the perception of different personalities and emotions can be difficult when the user personified an agent interfering with the movement of other agents. However, the values of perceptions increase when agents consider people to be Normal Life agents. Thus, we can say that these results refute $H0_2$. In addition, we found an educational level effect, in which people with complete high school perceived more different personalities and emotions than people with higher education. 

Regarding $H0_3$, we found relationships between perceiving interactions and perceiving different personalities and emotions. In most cases, these results had a low correlation and occurred in Scenario 3 (for men and people with complete high school). We also found a strong correlation when we analyzed data from people familiar with CG. This is an interesting result and refutes the hypothesis, as it means that the person who is familiar with CG tends to find a relationship between interactions (between agents) with different personalities and emotions. This makes sense, as people familiar with CG may be used to observing interactions between agents, personality traits, and emotions in simulations, in games, etc. Furthermore, taking into account that this result happened in Scenario 3, which was related to Normal Life, people familiar with CG may also be used to interact in virtual environments in which agents take into account the participant's presence, such as in games. \red{In relation to games, the behavior of characters that is closer to reality can improve the game experience. Taking into account our result, we should think that people can perceive extraversion in motion animations, for example, the perception of an extraverted character heading towards a group of friends.}

\section{Final Considerations}
\label{sec:conclusion}
In our research, the data collected indicates that one of the key factors in the perception of users is the kind of interaction they have with the virtual environment. Moreover, we found out that users tended to only perceive interactions and \red{extraversion personality trait} on the scenarios when they actively interacted with the agents. 

This paper has some limitations: firstly, we use only one simulation scenario, i.e., agents enter the environment and go to the goal, in all tested situations. Other experiments could enrich the tests and maybe the conclusions. Also, the number of agents could vary as to their appearance and animation. In addition, we could try to have more users to better sustain our hypotheses. For future work, we plan to model all of the OCEAN factors to influence the agents' geometric behavior. As discussed in this paper, the geometric factors are only perceived when the user is actively interacting with the agents, to remedy this we plan to add facial expressions for the agents to complement the geometric factors. Also, we plan to implement more than one physical appearance for the agents, so the realism can be increased. In addition, including more variety to the visual representation of the agent, that is, adding more 3D models of people, as to increase simulation diversity and realism, are part of our plans for the future.

\section*{ACKNOWLEDGMENT}

The authors would like to thank CNPq and CAPES for partially funding this work.

\bibliographystyle{IEEEtran}
\bibliography{bib}

\end{document}